\documentclass[conference]{IEEEtran}
\IEEEoverridecommandlockouts
\usepackage{cite}
\usepackage{amsmath,amssymb,amsfonts}
\usepackage{algorithmic}
\usepackage{graphicx}
\usepackage{textcomp}
\usepackage{xcolor}
\usepackage[normalem]{ulem}
\useunder{\uline}{\ul}{}

\def\BibTeX{{\rm B\kern-.05em{\sc i\kern-.025em b}\kern-.08em
    T\kern-.1667em\lower.7ex\hbox{E}\kern-.125emX}}
\begin{document}

\title{Agent vs. Avatar: \\Comparing Embodied Conversational Agents Concerning Characteristics of the Uncanny Valley}

\author{\IEEEauthorblockN{Markus Thaler}
\IEEEauthorblockA{\textit{Management, Communication \& IT} \\
\textit{MCI -- The Entrepreneurial School}\\
Innsbruck, Austria \\
mark.thaler@mci4me.at}
\and
\IEEEauthorblockN{Stephan Schlögl}
\IEEEauthorblockA{\textit{Management, Communication \& IT} \\
\textit{MCI -- The Entrepreneurial School}\\
Innsbruck, Austria \\
stephan.schloegl@mci.edu}
\and
\IEEEauthorblockN{Aleksander Groth}
\IEEEauthorblockA{\textit{Management, Communication \& IT} \\
\textit{MCI -- The Entrepreneurial School}\\
Innsbruck, Austria \\
aleksander.groth@mci.edu}
}

%\IEEEoverridecommandlockouts
%\IEEEpubid{\makebox[\columnwidth]{978-1-7281-5871-6/20/\$31.00~\copyright2020 IEEE %\hfill} \hspace{\columnsep}\makebox[\columnwidth]{ }}
%978-1-7281-5871-6/20/$31.00 ©2020 IEEE

\maketitle

\IEEEpubidadjcol

\begin{abstract}
Visual appearance is an important aspect influencing the perception and consequent acceptance of Embodied Conversational Agents (ECA). To this end, the Uncanny Valley theory contradicts the common assumption that increased humanization of characters leads to better acceptance. Rather, it shows that anthropomorphic behavior may trigger feelings of eeriness and rejection in people. The work presented in this paper explores whether four different autonomous ECAs, specifically build for a European research project, are affected by this effect, and how they compare to two slightly more realistically looking human-controlled, i.e. face-tracked, ECAs with respect to perceived humanness, eeriness, and attractiveness. Short videos of the ECAs in combination with a validated questionnaire were used to investigate potential differences. Results support existing theories highlighting that increased perceived humanness correlates with increased perceived eeriness. Furthermore, it was found, that neither the gender of survey participants, their age, nor the sex of the ECA influences this effect, and that female ECAs are perceived to be significantly more attractive than their male counterparts. %[150-250 words]
\end{abstract}


%=================================================================================================
\section{Introduction}
%=================================================================================================
For decades, research and industry have been struggling to build human-like computing systems, i.e. digital assistants, that can be operated via natural interaction modalities (e.g. natural language, gestures, mimics, etc.) and are capable of understanding and expressing emotions. 
%Until a few years ago, these systems were considered distant dreams of the future. 
Due to past developments in artificial intelligence and natural language processing, and the increasing uptake of digital assistants in the consumer sector, this gap between science fiction and reality is, however, gradually closing. With respect to application areas for these types of intelligent systems, research is often looking at the social and health care context, where people may benefit from support in various quotidian tasks~\cite{cereghetti2015virtual,fadhil2018patient,olaso2019empathic}. Here, the question as to which social characteristics a digital agent should or should not exhibit in these settings, is continuously being researched~\cite{dautenhahn1998TheAO,10.1145/3340764.3340793,sturgeon2019perception}. One of these issues concerns the right embodiment of respective systems, and how such may influence their credibility~\cite{duffy2003anthropomorphism,Kushmerick1997SoftwareAA,nowak2003anthropomorphism}. Humanization, in particular, may pose unwanted side-effects where a too realistic appearance is found to trigger feelings of eeriness in the human interlocutor -- a phenomenon that was already described in the early 20\textsuperscript{th} century by Jentsch, observing interactions with automatic dolls~\cite{Jentsch1906unheimlichkeit}. The heron built \textit{Uncanny Valley} effect demonstrates that the affinity with a given object is based on a non-linear relationship with the human-likeness of the respective object~\cite{mori2012uv}. Our study examines to which extent this effect can be found with the Embodied Conversational Agents (ECA) used in the European research project EMPATHIC\footnote{http://www.empathic-project.eu/}. In particular, our goal was to investigate the following research question: \\

\textit{How do human-driven, i.e. face-tracked, ECAs (so-called avatars) compare to autonomous ECAs with respect to their perceived humanness, eeriness and attractiveness?}

%=================================================================================================
\section{Related Work} 
\begin{figure}[ht]
	\centering
	\includegraphics[scale=0.6]{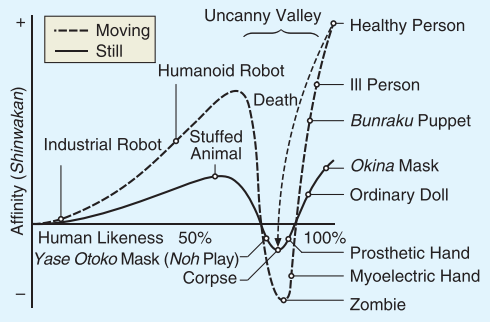} 
	\caption{The Uncanny Valley according to Mori~\cite{mori2012uv}}
	\label{img:mori-uv}
\end{figure}

%=================================================================================================
Although recent years have seen great progress in the development of speech and language-based user interfaces, people still feel uneasy when `talking' to an artificial entity~\cite{edu2002antAgents,Wei2015EvaluatingEC}. One reason for this may be rooted in the fact, that a large part of human communication lies outside the scope of pure language. ECAs aim to ameliorate this shortcoming through an additional emotional interaction channel. That is, they not only support speech-based interaction~\cite{Ciechanowski2018InTS,Wei2015EvaluatingEC}, but also feature some sort of visual and/or physical appearance~\cite{franklin1996agent}, which allows for the generation of facial expressions. Consequently, the human interlocutors' abilities to predict an agent's behaviour improve~\cite{koda1996agents}. In addition, the embodiment lets an agent appear more aesthetic and life-like~\cite{cassell2001ECA,dautenhahn1998TheAO} as well as potentially more intelligent~\cite{Kushmerick1997SoftwareAA}. Yet, improvement stops once the agent crosses the so-called \textit{Uncanny Valley} (cf. Fig.~\ref{img:mori-uv}). 

%=================================================================================================
\subsection{The Uncanny Valley} \label{uv-effect}
%=================================================================================================
According to Mori~\cite{mori2012uv}, the \textit{Uncanny Valley} (UV) shows the affinity a human observer has towards an object and how such changes when the human likeness of the object increases. Ho \& MacDorman~\cite{macdorman2010UVrevised} describe this as a graph illustrating a non-linear relationship between the human resemblance of an anthropomorphic entity and a viewer's emotional response. As can be seen in Fig.~\ref{img:mori-uv}, Mori chose industrial robots as a starting point, since they exhibit little resemblance to humans. At the opposite side of the valley one then finds for example realistic hand prostheses, as they are rather difficult to distinguish from real human hands. According to Seyama \& Nagayama~\cite{seyama2007UVrealism}, the UV effect can be shown with both physical and virtual objects. In an experiment with computer-generated characters MacDorman and colleagues, for example, found that faces with photo-realistic textures and a high degree of detail tend to trigger eeriness. Characters with a lower photo-realism level (i.e., 75\%), however, were deemed the least uncanny by human viewers~\cite{macdorman2009UncResponses}. Hence, it may be argued that the UV theory contradicts the view that a robot or embodied agent should resemble a human being as close as possible~\cite{dautenhahn1998TheAO}.
 
The second part of the UV theory deals with motion. According to Mori~\cite{mori2012uv}, the amplitude of the curve increases when the object is in motion. In an experiment with avatars (i.e. human-controlled agents), it was found that motion influences familiarity and perceived attractiveness of the agent~\cite{McDonnell2012RenderStyle}. Although, it has to be underlined that both familiarity and perceived attractiveness are subject to personal taste. For example, it was shown that even very human-like characters may be deemed repelling~\cite{seyama2007UVrealism}. Similarly, Jentsch notes in his definition of uncannyness, that a realistic technological stimulus can trigger a feeling of eeriness in people~\cite{Jentsch1906unheimlichkeit}. There are several theories where this sensation of eeriness might come from~\cite{seyama2007UVrealism}. Bartneck et al. attribute it to the framing theory~\cite{bartneck2007UVcliff}, which states that when observing new things, humans refer to past experiences, so-called frames. This gives rise to expectations that may not always be met. With human-like robots, for example, the `human frame' is selected, but when seeing the robot's mechanical movements, which may resemble those of a sick or injured person, it is often the case that a feeling of discomfort arises. Another trigger for this eeriness may be found in the skin discoloration of artificial characters, which reminds people of death and thus may evoke respective fears~\cite{burleigh2013UVexist, Ciechanowski2018InTS,macdorman2006subjective}. Here one might also draw a connection to the psychological process of anthropomorphism.

%=================================================================================================
\subsection{Anthropomorphism} \label{anthropomorphism}
%=================================================================================================
The anthropomorphism phenomenon is one crucial aspect related to the UV theory, stating that human characteristics, motivations, intentions and emotions may be attributed to non-human entities, even if these have only slightly human-like traits~\cite{epley2007anthropomorphism}. Duffy~\cite[p.~180]{duffy2003anthropomorphism} describes this as \textit{``attributing cognitive or emotional states to something based on observation in order to rationalise an entity's behaviour in a given social environment''}. In other words, anthropomorphism is considered a psychological process helping humans rationalize observations. Guthrie~\cite{guthrie1980religion} describes the respective procedure in four consecutive steps: (1) one is confronted with an unknown object which initially triggers shallow conclusions; (2) this initial conclusions are then combined with and expanded upon past experiences; (3) the most probable template describing the object is selected; and (4) since the `human' template is most commonly chosen, the observed characteristics are compared to those of humans. As this process is based on non-rational thinking, humanization can be perceived negatively~\cite{tondu2012anthropomorphism}. Contrary to this, in Human-Computer Interaction research and design, anthropomorphism is often considered a design guideline that helps build appealing user interfaces (UI), and aims at triggering rather pleasant experiences in users~\cite{dautenhahn1998TheAO}. While humans often anthropomorphise, even in cases where an interface exhibits hardly any human traits, it has been shown that the context and given task setting influence the perception and consequent acceptance of this human-likeness~\cite{fong2003social}. In social tasks, for example, a human-like agent seems acceptable, whereas in other settings an anthropomorphic appearance may be less expected, and consequently trigger feelings of uneasiness~\cite{ring2014agent}. 

%=================================================================================================
\subsection{ECAs in Social and Health Care Settings}
%=================================================================================================
ECAs may help simplify human-technology interaction. For seniors, in particular, accessibility to technology can be improved ~\cite{peeters2017ECA, olaso2019empathic}. One area  that sees an ongoing uptake of these types of intelligent systems is the social and health care sector, where ECAs are increasingly used to support everyday tasks of people~\cite{micoulaud2016acceptability}. However, as already outlined above, the visual appearance of these ECAs poses certain challenges with respect to their human-likeness and the consequent effects on people's feelings of eeriness. Thus, the goal of our work was to explore how  ECAs specifically created for health related tasks, i.e. the ECAs built for the EMPATHIC project, are perceived by the general population. 

%=================================================================================================
\section{Methodology} \label{sec:methodology}
%=================================================================================================
\begin{figure}
	\centering
	\includegraphics[scale=0.16]{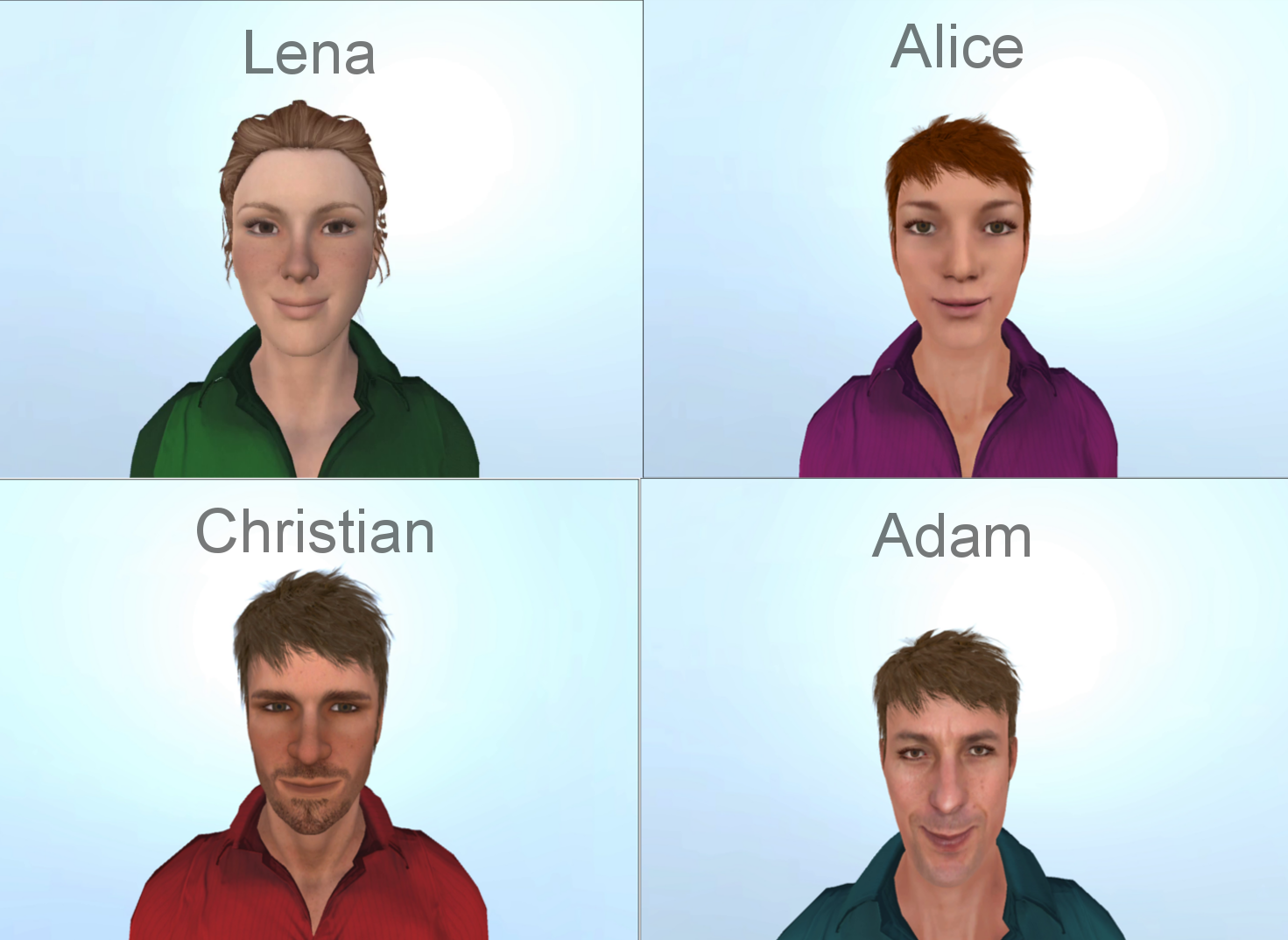} 
	\caption{Autonomous ECAs built for the EMPATHIC project}
	\label{img:agents}
\end{figure}
\begin{figure}
	\centering
	\includegraphics[scale=0.25]{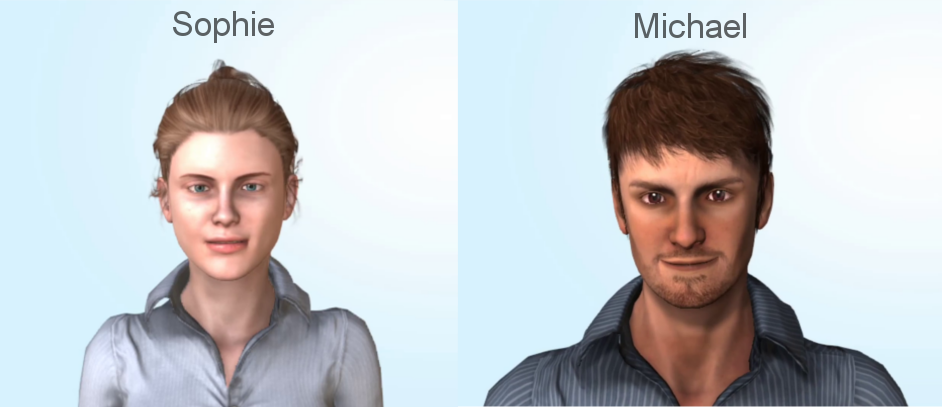} 
	\caption{Human-driven ECAs (built using the Reallusion Software Suite)}
	\label{img:avatars}
\end{figure}

The goal was to evaluate six different ECAs (cf. Figures \ref{img:agents} \& \ref{img:avatars}). Four autonomous agents, which were created for the EMPATHIC project, i.e. Lena, Alice, Christian and Adam, and two human-driven/face-tracked agents (so-called avatars), i.e. Sophie and Michael, which were built separately, so as to serve as a control group (note: for these two agents we used the Reallusion Software Suite\footnote{https://www.reallusion.com/iclone/}, which led to slightly more realistic faces and facial expressions than what was achieved by the project ECAs). All six ECAs exhibited similar characteristics with respect to Ring et al.'s classification of anthropomorphic UIs~\cite{ring2014agent}. That is, all were human(-like), mimicking the same ethnicity and similar age and wearing similar clothing. Only the level of realism distinguished the project ECAs (i.e., Lena, Alice, Christian and Adam) from the slightly more realistic Reallusion ECAs (i.e., Sophie and Michael). In accordance with the above discussed literature we thus formed the following hypotheses to be tested:
\begin{itemize}
    \item H1: Human-driven ECAs (i.e., avatars) exhibit a higher level of perceived humanness than autonomous ECAs (cf. MacDorman et al.~\cite{macdorman2009UncResponses}).
    \item H2: The perceived level of humanness has a significant influence on the perceived level of eeriness (cf. McDonnell et al.~\cite{McDonnell2012RenderStyle}).
    \item H3a: There is a significant difference between genders when it comes to the perceived level of eeriness (cf. Tinwell \& Sloan~\cite{TINWELL2014286}).
    \item H3b: There is a significant difference between genders when it comes to the perceived level of attractiveness (cf. Tinwell \& Sloan~\cite{TINWELL2014286}).
    \item H4a: There is a significant difference with respect to perceived eeriness dependent on the sex of the ECA (cf. Esposito et al.~\cite{esposito2018seniors}).
    \item H4b: There is a significant difference with respect to perceived attractiveness dependent on the sex of the ECA (cf. Esposito et al.~\cite{esposito2018seniors}).
    \item H5a: There is a significant positive correlation between a participant's age and the perceived eeriness of ECAs (cf. Yaghoubzadeh et al.~\cite{yaghoubzadeh2012toward}).
    \item H5b: There is a significant negative correlation between a participant's age and the perceived attractiveness of ECAs (cf. Yaghoubzadeh et al.~\cite{yaghoubzadeh2012toward}).
    \item H6: Human-driven ECAs (i.e., avatars) exhibit a higher level of perceived eeriness than autonomous ECAs.
\end{itemize}

In order to evaluate these hypotheses and consequently the extent to which the tested ECAs may fall into the UV, we exposed participants to an eight-second video of an agent and subsequently asked them to rate the agent's perceived level of humanness, eeriness and attractiveness according to the following 21 UV semantic differential effect scales proposed by MacDorman \& Ho~\cite{ho2017measureUV} (note: in brackets we provide the German translation of the terms the way they were used in the questionnaire):
\begin{itemize}
    \item \textbf{Humanness:} 7-point semantic differentials on
    \begin{itemize}
        \item inanimate $\leftrightarrow$ living (ge: unbelebt/lebendig)
        \item synthetic $\leftrightarrow$ real (ge: synthetisch/echt)
        \item mechanical movement $\leftrightarrow$ biological movement \\ (ge: mech. Bewegungen/nat. Bewegungen)
        \item human-made $\leftrightarrow$ humanlike \\ (ge: von Menschen gemacht/menschen\"{a}hnlich) 
        \item without definite lifespan $\leftrightarrow$ mortal \\ (ge: ohne definitive Lebensdauer/sterblich)
        \item artificial $\leftrightarrow$ natural (ge: k\"{u}nstlich/natürlich)
    \end{itemize}
    \item \textbf{Eeriness:} 7-point semantic differentials on
    \begin{itemize}
        \item dull $\leftrightarrow$ freaky (ge: eintönig/ausgefallen)  
        \item predictable $\leftrightarrow$ eerie (ge: absch\"{a}tzbar/unheimlich)  
        \item plain $\leftrightarrow$ weird (ge: schlicht/sonderbar)                           
        \item ordinary $\leftrightarrow$ supernatural \\ (ge: gew\"{o}hnlich/außernat\"{u}rlich)                   
        \item boring $\leftrightarrow$ shocking (ge: langweilig/schockierend)                
        \item uninspiring $\leftrightarrow$ spine-tingling \\ (ge: uninspirierend/elektrisierend)             
        \item predictable $\leftrightarrow$ thrilling (ge: vorhersehbar/mitrei{\ss}end)                    
        \item bland $\leftrightarrow$ uncanny (ge: nichtssagend/untypisch)                       
        \item unemotional $\leftrightarrow$ hair-raising \\ (ge: emotionslos/furchterregend)                  
        \item  familiar $\leftrightarrow$ uncanny (ge: vertraut/unheimlich)                       
    \end{itemize}
    \item \textbf{Attractiveness:} 7-point semantic differentials on
    \begin{itemize}
        \item ugly $\leftrightarrow$ beautiful (ge: h\"{a}sslich/sch\"{o}n)                              
        \item repulsive $\leftrightarrow$ agreeable (ge: absto{\ss}end/ansprechend)                       
        \item crude $\leftrightarrow$ stylish (ge: geschmackslos/stylisch)                       
        \item messy $\leftrightarrow$ sleek (ge: unordentlich/gepflegt)                       
        \item unattractive $\leftrightarrow$ attractive (ge: unattraktiv/attraktiv)                     
    \end{itemize}
\end{itemize}

This procedure was repeated for each of the six ECAs. Finally, participants were asked to provide some additional demographic information. We particularly targeted German-speaking participants so as to prevent any potential cultural bias. To this end, all scales and questions were translated. A German-speaking lecturer of English as well as a bilingual colleague were consulted to improve translation quality. In addition, all scales were tested for reliability before further evaluations began. Participation in the study was voluntary (note: three EUR 10,- Amazon Gift Vouchers were raffled off among the participants). Furthermore, questionnaire and respective procedure were evaluated by MCI's research ethics group for accordance with ethical guidelines concerning research with human participation.

%=================================================================================================
\section{Results}
%=================================================================================================
A total of $n=215$ participants (150 females) completed the above described procedure, approx. 50\% (i.e. 108) of whom were students (predominantly business and information systems but also other fields such as pharmacy and physics). The overall age distribution ($74\%\leq30~years , 21\%~ 31-60~years, 5\%>60~years$) shows a strong right-bound skewness of 1.766 ($Mean=31.04; Median=25$). 
Scale reliability for humanness, eeriness and attractiveness were first tested individually for each stimulus. Then each dimension was analyzed in its entirety using Cronbach's $\alpha$. Table~\ref{tab:reliability} shows that the overall scale reliability was excellent ($\alpha=0.95$) and when focusing on internal stimuli only, it was still good ($\alpha>0.70$)~\cite{ho2017measureUV}. Given this reliability of scales we were able to proceed, evaluating and consequently comparing the six ECAs with respect to their perceived humanness, eeriness, and attractiveness. Respective results are summarized in the following sub-sections.

\begin{table}
\centering
\caption{Reliability of Scales}
\begin{tabular}{lccc}
\hline
\textbf{Stimulus}   & \textbf{Humanness ($\alpha$)} & \textbf{Attractiveness ($\alpha$)} & \textbf{Eeriness ($\alpha$)} \\
\hline
Lena & 0.871 & 0.896 & 0.827 \\
Adam & 0.925 & 0.896 & 0.856 \\
Sophie & 0.937 & 0.932 & 0.864 \\
Christian & 0.956 & 0.910 & 0.878 \\
Alice & 0.948 & 0.921 & 0.878 \\
Michael & 0.947 & 0.951 & 0.894 \\
\hline
\textbf{Overall} & 0.957 & 0.926 & 0.950 \\  
\hline
\end{tabular}
\label{tab:reliability}
\end{table}

%=================================================================================================
\subsection{Humanness}
\begin{figure}[ht]
	\centering
	\includegraphics[scale=0.35]{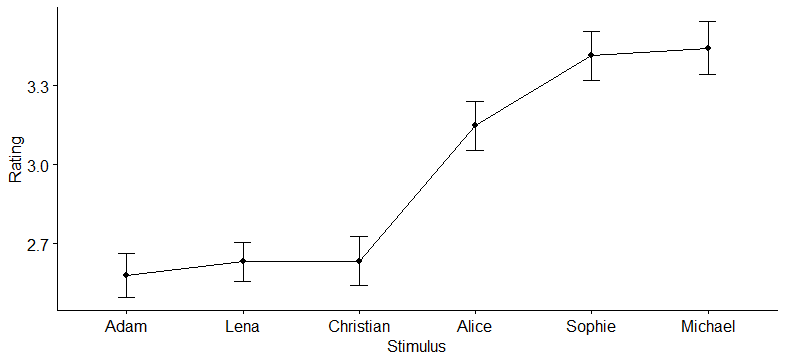} 
	\caption{Mean Values of Perceived Humanness for Each of the Shown ECAs}
	\label{img:mean-humanness-stimuli}
\end{figure}

%=================================================================================================
Fig.~\ref{img:mean-humanness-stimuli} shows the mean values of the six ECAs with respect to humanness (note: we calculated the mean over all the scales concerning humanness described in Section~\ref{sec:methodology} where values close to 1 signify low humanness and values close to 7 signify high humanness). To this end, the project ECAs Adam ($\Bar{x}=2.58; SD=1.22$), Lena ($\Bar{x}=2.63; SD=1.09$) and Christian ($\Bar{x}=2.63; SD=1.36$) were perceived quite similarly. Alice, however, was perceived as more human-like ($\Bar{x}=3.15; SD=1.39$). The two Reallusion ECAs Sophie ($\Bar{x}=3.41; SD=1.38$) and Michael ($\Bar{x}=3.44; SD=1.48$) were also perceived similarly and more human-like than any of the project ECAs. Yet, Michael's ratings showed the highest variability. Although, human similarity exhibited generally a higher standard deviation than the other dimensions.%($\Bar{x}_(over all stimuli)$=2.98; SD=0.96). 

A T-test for combined samples between the most human-like project ECA (i.e., Alice) and the least human-like Reallusion ECA (i.e., Sophie) points to a significant difference in perceived humanness ($p=0.014$) attributed to the type of control, i.e. autonomous vs. human-controlled. An ANOVA including the Scheff\'{e} post-hoc test generally confirmed this significantly higher level of perceived humanness with the Reallusion ECAs, yet further showed that also Alice (the most human-like project ECA) scored significantly higher than the other project ECAs (cf. Table~\ref{tab:anova}). All in all, however, we may argue that our data supports H1. 

\begin{table}
\centering
\caption{ANOVA - Post-Hoc Scheff\'{e}}
\begin{tabular}{lll}
\hline
\textbf{Stimuli} & \textbf{Humanness ($p$)} & \textbf{Eeriness ($p$)} \\
\hline
\textbf{Alice} $\longleftrightarrow$ Adam  & 0.0015**  & 0.8295    \\
Christian $\longleftrightarrow$ Adam & 0.9994 & 0.9988    \\
Lena $\longleftrightarrow$ Adam & 0.9995 & 0.9976    \\
\textbf{Michael} $\longleftrightarrow$ Adam & 0.0000*** & 0.0000*** \\
\textbf{Sophie} $\longleftrightarrow$ Adam & 0.0000*** & 0.0903 \\
Christian $\longleftrightarrow$ \textbf{Alice} & 0.0066**  & 0.5834    \\
Lena $\longleftrightarrow$ \textbf{Alice} & 0.0064** & 0.5441    \\
\textbf{Michael} $\longleftrightarrow$ Alice & 0.3798 & 0.0068**  \\
Sophie $\longleftrightarrow$ Alice & 0.5038 & 0.7548    \\
Lena $\longleftrightarrow$ Christian & 1.0000 & 1.0000    \\
\textbf{Michael} $\longleftrightarrow$ Christian & 0.0000*** & 0.0000*** \\
\textbf{Sophie} $\longleftrightarrow$ Christian & 0.0000*** & 0.0266*  \\
\textbf{Michael} $\longleftrightarrow$ Lena & 0.0000*** & 0.0000*** \\
\textbf{Sophie} $\longleftrightarrow$ Lena & 0.0000*** & 0.0219*  \\
Sophie $\longleftrightarrow$ Michael & 1.0000 & 0.3367 \\
\hline
\end{tabular}
\label{tab:anova}
\end{table}

%=================================================================================================
\subsection{Eeriness}
\begin{figure}[ht]
	\centering
	\includegraphics[scale=0.35]{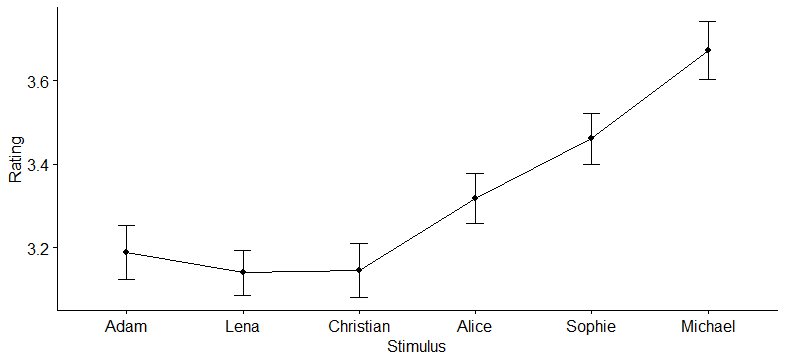} 
	\caption{Mean Values of Perceived Eeriness for Each of the Shown ECAs}
	\label{img:mean-eeriness-stimuli}
\end{figure}

%=================================================================================================
In order to explore the UV effect, ECAs were ordered according to their human similarity scores and then evaluated with respect to their perceived level of eeriness. Fig.~\ref{img:mean-eeriness-stimuli} shows that after initial improvements eeriness starts increasing with increasing humanness. That is, the project ECAs Adam ($\Bar{x}=3.19; SD=0.94$), Lena ($\Bar{x}=3.14; SD=0.80$) and Christian ($\Bar{x}=3.15; SD=0.95$) had lower eeriness scores than Alice ($\Bar{x}=3.32; SD=0.89$). Similarly, the Reallusion ECAs Sophie ($\Bar{x}=3.46; SD=0.90$) and Michael ($\Bar{x}=3.67; SD=1.01$) scored rather negatively, with Michael moving even further away from Sophie % (difference $\delta \Bar{x}$=0.21) 
and again showing the greatest variability in people's perception of eeriness. A T-test for independent samples showed that this perception was independent of people's gender (F-test for gender variance equality: $p=0.752$), for which H3a had to be rejected ($p=0.655$). Similarly with respect to the agents' sex no significant difference in perceived eeriness was found ($p=0.411$). Hence, also H4a was rejected.

Looking at a connection between humanness and eeriness, a Pearson correlation analysis points to a medium to strong positive correlation between those two variables ($r=0.573; p=0.000$). The subsequently performed linear regression shows that perceived humanness is able to explain 32.9\% of the variance in perceived eeriness ($R^2=0.329; Beta=0.573; p=0.000$). Hence, H2 is supported by the data.

With respect to H5a, we explored a potential positive relation between a participant's age and the ECA's perceived eeriness. The performed Pearson correlation analysis, however, shows a weak but significant negative connection between those two variables ($r=-0.150; p=0.014$). Consequently, H5a had to be rejected. 

Finally, a T-test for combined samples pointed to a significant difference between the most eerie project ECA (i.e., Alice) and the least eerie Reallusion ECA (i.e., Sophie), supporting the assumption outlined by H6 ($p=0.024$). The subsequently performed ANOVA and Scheff\'{e} post-hoc test confirmed this significantly higher level of perceived eeriness for Michael (i.e., significantly higher eeriness scores than all project ECAs) and partly for Sophie (i.e., significantly higher eeriness scores than Christian and Lena). Overall, we thus may argue that our data supports H6 in that the perceived eeriness of the two Reallusion ECAs (i.e., the avatars) was rated significantly higher than the perceived eeriness of the four project ECAs (cf. Table~\ref{tab:anova}).

%=================================================================================================
\subsection{Attractiveness}
\begin{figure}[ht]
	\centering
	\includegraphics[scale=0.35]{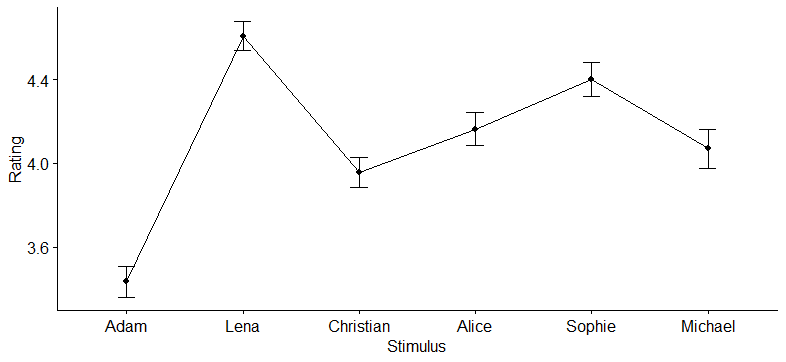} 
	\caption{Mean Values of Perceived Attractiveness for Each of the Shown ECAs}
	\label{img:mean-attractiveness-stimuli}
\end{figure}

%=================================================================================================
Attractiveness was generally rated higher than the other two dimensions. Lena ($\Bar{x}=4.61; SD=1.01$) and Sophie ($\Bar{x}=4.40; SD=1.21$) were rated the most attractive, Adam the least attractive ($\Bar{x}=3.43; SD=1.07$), and Alice ($\Bar{x}=4.16; SD=1.17$), Michael ($\Bar{x}=4.07; SD=1.35$) and Christian ($\Bar{x}=3.96; SD=1.08$) lay somewhere in between. Also here the participants agreed the least with the evaluation of Michael. In Fig.~\ref{img:mean-attractiveness-stimuli} the ECAs were ranked according to their perceived humanness. Looking at the graph, it is noticeable that the female agents were consistently classified as more attractive. A T-test on the mean values of the ECA's sex supports this assumption, showing a significant difference in terms of perceived attractiveness ($p=0.000$). Consequently, H4b is supported by the data. From a participants' point of view, however, no gender differences with respect to the perceived attractiveness of ECAs was found ($p=0.278$). Hence H3b had to be rejected. 

Finally, investigating a potential negative relation between the participants' age and the perceived attractiveness of ECAs, no significant correlation was found ($r= -0.095; p=0.164$), for which also H5b had to be rejected. 

%=================================================================================================
\section{Summary, Limitations and Outlook}
%=================================================================================================

The goal of our study was to examine six different ECAs concerning characteristics of the UV. According to the classification by Ring and colleagues~\cite{ring2014agent} we expected a more realistic rendering style to perform better than a less realistic rendering style. Consequently, we evaluated the ECAs with respect to their perceived humanness, attractiveness and eeriness. The results of our evaluation do not fully confirm our expectations, yet they are in line with previous work. That is, we found a positive correlation between an ECA's human similarity and its perceived eeriness. Analogous to the UV theory, this means that with increasing realism, the viewer's affinity with an ECA decreases. 
Age-related connections could only be confirmed to a limited extent, showing a slight negative correlation with the participant's eerieness ratings. The participant's gender had no effects on the ratings, the sex of the stimuli, however, influenced their perceived attractiveness, with female agents being perceived significantly more attractive than male ones. 

%Limitations
Although these results generally confirm previous work, we would like to point to a number of limitations of our study. First, while our human-driven ECAs were indeed perceived more human-like than the autonomous ECAs built for the EMPATHIC project, their level of humanness was still rather low (i.e., $\Bar{x}=3.41$ and $\Bar{x}=3.44$ respectively). Such may have had an affect on the significance of results, in particular with respect to the origin of perceived eeriness. Second, the used questionnaire was rather long and monotonous, repeating the same sections (i.e. MacDorman \& Ho's 21 semantic differential effect scales) six times, which may have influenced people's responses. Third, the negative correlation we found with respect to  participants' age and their provided eeriness ratings was very weak (cf. H5a). Finally, a more homogeneous sample may have potentially provided better insights. That is, although in general our data did not point to any irregularities, the demographic distribution was not ideal. The majority of respondents were students younger than 30 years, and more than two-thirds were female. 

%Outlook
In conclusion, this study has confirmed previous findings with respect to the UV theory; i.e. the more human-like an ECA appears the greater its produced feeling of eeriness. An additional, not negligible result of our study, is the translation and validation of MacDorman \& Ho's UV scales into German, which may allow for additional future studies in German-speaking countries. Those studies may, for example, focus on the level of realism that is accepted with ECAs. To this end an interesting question to investigate may be whether ECAs could be made so realistic and credible that they would `climb up' the other end of the UV. Also, it may be worth exploring whether female ECAs are consistently rated more attractive than male ones. Finally, concerns regarding privacy and ethics connected to the use of ECAs would require additional research. In particular, if their potential application domain is focusing on social and health care settings.

%=================================================================================================
\section*{Acknowledgment}
%=================================================================================================
The research presented in this paper is conducted as part of the project EMPATHIC that has received funding from the European Union's Horizon 2020 research and innovation programme under grant agreement No 769872.

\bibliography{ichms}
\bibliographystyle{IEEEtran}

\end{document}